# Brownian Emitters

Roumen Tsekov

Department of Physical Chemistry, University of Sofia, 1164 Sofia, Bulgaria

A Brownian harmonic oscillator, which dissipates energy either by friction or via emission of electromagnetic radiation, is considered. This Brownian emitter is driven by the surrounding thermo-quantum fluctuations, which are theoretically described by the fluctuation-dissipation theorem. It is shown how the Abraham-Lorentz force leads to dependence of the half-width on the peak frequency of the oscillator amplitude spectral density. It is also found that for the case of a charged particle, moving in vacuum at zero temperature, its root-mean-square velocity fluctuation is a universal constant, equal to roughly 1/18 of the speed of light. The relevant Klein-Kramers and Smoluchowski equations are derived as well.

The Brownian harmonic oscillator is a very well-known and extensively studied theoretical model. It describes particularly the thermal vibrations of molecules in chemistry, which are the core of the IR spectroscopy. In the present paper the standard model of a Brownian oscillator is extended by taking into account another way of energy dissipation via emission of electromagnetic radiation. Thus, the Langevin equation for a charged harmonic oscillator acquires the form

$$\ddot{X} + \gamma_0 \dot{X} + \omega_0^2 X = F/m + \tau_0 \dddot{X} \tag{1}$$

The second term on the left hand-side describes standard liner friction by the environment with a specific friction constant $\gamma_0$, while $F$ is the stochastic Langevin force. Since the oscillator is a vibrating dipole own frequency $\omega_0$, it emits permanently electromagnetic radiation according to electrodynamics. This effect is described in Eq. (1) by the last term, being the acceleration due to the Abraham-Lorentz force [1]. The characteristic time-constant $\tau_0 \equiv e^2/6\pi\varepsilon_0 mc^3$ is very small and for electrons, for instance, it corresponds to a characteristic frequency $2\pi/\tau_0$ of the order of a yottahertz (1 YHz = $10^{24}$ Hz). The Abraham-Lorentz force contradicts to Newtonian mechanics, which assumes that forces can depend on the particle position and velocity only. Furthermore, its relativistic character is emphasized by the presence of the speed of light $c$ in $\tau_0$. Equation (1) is a Langevin equation with two friction mechanics and it can be easily extended to an arbitrary external potential. In the case of radiation only, Eq. (1) is proposed as the core of stochastic electrodynamics [2], which is looking for an explanation for the origin of quantum mechanics [3]. Puthoff [4] has derived the ground state of hydrogen on the base of a frictionless Eq. (1), assuming that the Langevin force represents random radiation of the zero-point fluctuation energy in vacuum. A relativistic extension of Eq. (1) is also proposed [5]. In the treatments above the fluctuation force is described from the viewpoint of electrodynamics. Hence, it is not recognized as the Langevin force, subject of the fluctuation-dissipation theorem.

Applying the standard Fourier transformation, Eq. (1) reduces straightforward to

$$X_\omega = \frac{F_\omega / m}{-\omega^2 + i\omega\gamma_\omega + \omega_0^2} \qquad (2)$$

where the frequency-dependent friction coefficient $\gamma_\omega \equiv \gamma_0 + \omega^2 \tau_0$ indicates a non-Markovian character and memory effects [6]. Multiplying Eq. (2) by its complex-conjugated one and taking a statistical average yields an expression for the oscillator amplitude spectral density

$$S_{XX} = \frac{S_{FF}/m^2}{(\omega^2 - \omega_0^2)^2 + \omega^2 \gamma_\omega^2} \qquad (3)$$

The fluctuation-dissipation theorem $S_{FF} = m\gamma_\omega \hbar\omega \coth(\hbar\omega/2k_BT)$ relates the Langevin force spectral density to the overall friction coefficient [6]. Note that the radiative contribution here contains either the zero-point fluctuation energy in vacuum or the Planck spectrum of black-body electromagnetic radiation. Substituting $S_{FF}$ in Eq. (3) results in the final expression for the oscillator spectral density in the form

$$S_{XX} = \frac{\hbar \coth(\hbar\omega/2k_BT)}{2m\omega} \frac{2\omega^2 \gamma_\omega}{(\omega^2 - \omega_0^2)^2 + \omega^2 \gamma_\omega^2} \qquad (4)$$

The first term represents the amplitude dispersion for an oscillator with an arbitrary frequency. Note that the reduced Planck constant $\hbar$ appears solely from the Langevin force spectral density, i.e. Eq. (1) describes a classical particle moving in a quantum environment. The expression (4) decreases asymptotically at large frequencies inversely proportional to the third power of $\omega$. This means that the integral of the spectral density $S_{VV} = \omega^2 S_{XX}$ of the oscillator velocity fluctuations diverges logarithmically, which indicates the existence of an upper cut-off limit. Indeed, combining the time-energy Heisenberg relation $\Delta t \Delta E \geq \hbar/2$ with the Einstein relativity one $\Delta E \leq mc^2$ yields a cut-off $\Omega \equiv 2mc^2/\hbar$, being to the Zitterbewegung frequency [3]. Surprisingly, the dimensionless product $\Omega \tau_0 = 4\alpha/3 \approx 1/103$ is a universal small number, where $\alpha \approx 1/137$ is the fine-structure constant.

At low friction the second term in Eq. (4) exhibits a sharp maximum at $\omega = \omega_0$. For this reason, one can apply the so-called relaxation approximation, replacing everywhere $\omega$ by the resonant frequency $\omega_0$ but keeping their difference $\omega - \omega_0$ constant. Thus, Eq. (4) reduces to

$$S_{XX} = \frac{\hbar \coth(\hbar \omega_0 / 2k_B T)}{2m\omega_0} \frac{\gamma_{\omega_0}/2}{(\omega - \omega_0)^2 + \gamma_{\omega_0}^2/4} \tag{5}$$

The first term here represents the oscillator dispersion, which is in accordance to quantum statistical mechanics, while the second term is the Lorentzian function, well-known in spectroscopy. The spectral density (5) describes a resonant maximum at $\omega_0$ with half-width frequency given by $\gamma_{\omega_0} = \gamma_0 + \omega_0^2 \tau_0$. Therefore, due to the Abraham-Lorenz force, the width of the spectral peak becomes dependent on the peak frequency $\omega_0$. Because of the extremely low value of the characteristic time $\tau_0$, however, this effect is essential at very high peak frequency and very low friction constant $\gamma_0$. Note that the latter could also depend on $\omega_0$. For instance, in the case of a vibrating particle adsorbed on a solid surface, the friction constant is $\gamma_0 = 3\pi m \omega_0^4 / 2M\omega_D^3$, which depends also on the Debye cut-off frequency $\omega_D$ of the phonon spectrum in the solid [7].

According to Eq. (4) the spectral density of the oscillator velocity fluctuations reads

$$S_{VV} = \omega^2 S_{XX} = \frac{\hbar \omega \coth(\hbar \omega / 2k_B T)}{2m} \frac{2\omega^2 \gamma_\omega}{(\omega^2 - \omega_0^2)^2 + \omega^2 \gamma_\omega^2} \tag{6}$$

Equation (6) reduces to one for a free Brownian particle in the case of zero oscillator own frequency ($\omega_0 \equiv 0$)

$$S_{VV} = \frac{\hbar\omega\coth(\hbar\omega/2k_BT)}{2m} \frac{2\gamma_\omega}{\omega^2+\gamma_\omega^2} \tag{7}$$

It simplifies further in the classical limit $\hbar \to 0$ to

$$S_{VV} = \frac{k_BT}{m} \frac{2\gamma_\omega}{\omega^2+\gamma_\omega^2} \tag{8}$$

Neglecting the effect of the electromagnetic radiation ($\tau_0 \equiv 0$), the velocity autocorrelation function corresponding to Eq. (8) is the classical exponent $C_{VV} = (k_BT/m)\exp(-\gamma_0\tau)$, which provides the standard Einstein diffusion constant $D = S_{VV}(0)/2 = k_BT/m\gamma_0$. If the particle is moving in vacuum ($\gamma_0 \equiv 0$), the velocity autocorrelation function is exponential $C_{VV} = (k_BT/m)\exp(-\tau/\tau_0)$ again. However, the corresponding new diffusion constant $D_0 \equiv k_BT\tau_0/m$ is very low, since $\tau_0$ is extremely small. According the fluctuation-dissipation theorem, the classical Langevin force spectral density is $S_{FF} = 2mk_BT\gamma_\omega$. It is possible to invert analytically this Fourier image and the resulting autocorrelation function reads $C_{FF} = 2mk_BT(\gamma_0 - \tau_0\partial_\tau^2)\delta(\tau)$. Therefore, the classical Langevin force is a superposition of a standard white noise and another violet noise due to thermal Rayleigh-Jeans irradiation from the environment. It is also amazing that Eq. (1) for a free particle in vacuum can be integrated on time to obtain $-\ddot{X} + \dot{X}/\tau_0 = f/m$, where the Langevin force $f \equiv \int Fdt/\tau_0$ possesses the Ohmic spectral density $S_{ff} = (m/\tau_0)\hbar\omega\coth(\hbar\omega/2k_BT)$. This Langevin equation describes a particle with a negative mass such as an electron in the Dirac sea. Neglecting at large time the first term due to the very low value of $\tau_0$, the equation above reduces to a standard stochastic differential equation $m\dot{X}/\tau_0 = f$. In the classical limit the Langevin force is a white noise and $S_{ff} = 2mk_BT/\tau_0$ corresponds to a diffusion equation for the particle position probability density with diffusion constant $D_0$. Note that due to the radiation and very low value of $\tau_0$, the vacuum behaves as a very dissipative environment.

Perhaps, the most interesting case is the particle motion in vacuum ($\gamma_0 \equiv 0$) at zero temperature ($T \equiv 0$). In this purely quantum case Eq. (7) reduces to

$$S_{VV} = \frac{\hbar}{m} \frac{\omega \tau_0}{1 + \omega^2 \tau_0^2} \qquad (9)$$

The corresponding velocity autocorrelation function $C_{VV}$ is a unique function of the dimensionless time $\tau / \tau_0$. It is easy to calculate the velocity dispersion via

$$C_{VV}(0) = \frac{1}{\pi} \int_0^\Omega S_{VV} d\omega = \frac{\hbar}{2\pi m \tau_0} \ln(1 + \Omega^2 \tau_0^2) \approx \frac{\hbar \Omega^2 \tau_0}{2\pi m} = \frac{\Omega \tau_0}{\pi} \frac{\hbar \Omega}{2m} = \frac{4\alpha}{3\pi} c^2 \qquad (10)$$

Hence, the root-mean-square velocity fluctuation of a stochastically radiating charged particle, being in equilibrium with the vacuum zero-point quantum electromagnetic fluctuations, does not depend on the particle mass $m$ in contrast to the classical case. It is a universal constant, equal to roughly 1/18 of the speed of light $c$. Thus, the quantum particle trembling motion is an order of magnitude faster than that of 1s orbital electron in the hydrogen atom. Since $\omega \tau_0 \leq \Omega \tau_0 < 0.01$ one can simplify further Eq. (9) to $S_{VV} = \hbar \omega \tau_0 / m$, which shows that $S_{VV} / \omega^2$ describes a Flicker noise. Integrating properly the latter one can derive the position dispersion of the quantum particle $\sigma^2 = (2\hbar \tau_0 / \pi m) \ln(t / \tau_0)$, which increases logarithmically at large time. The characteristic length $(2\hbar \tau_0 / \pi m)^{1/2} = (4\alpha / 3\pi)^{1/2} (\hbar / mc)$ is about 1/18 of the Compton wavelength and for an electron, for instance, this mean-free path is 21 fm or roughly 4 electron diameters $2\alpha \hbar / mc$ [8]. Therefore, the frequency of collisions between the particle and virtual photons, being the electromagnetic field fluctuations in vacuum, is the Compton frequency $mc^2 / \hbar = \Omega / 2$.

It is interesting to see how probability equations look like for a Brownian emitter. Considering an arbitrary potential $U$ one can approximate the particle acceleration in the Abraham-Lorentz force by the Newtonian one $-\nabla U / m$. Thus, the extension of Eq. (1) reads

$$m\ddot{R} + m\gamma_0 \dot{R} + \nabla U = F - \tau_0 \nabla \dot{U} = F - \tau_0 \dot{R} \cdot \nabla \otimes \nabla U \qquad (11)$$

The Klein-Kramers equation for the phase-space distribution density $W(v,r,t)$ follows in the classical limit directly from Eq. (11) [9]

$$\partial_t W + v \cdot \nabla W - \nabla U \cdot \partial_v W / m = (\gamma_0 I + \tau_0 \nabla \otimes \nabla U / m) : \partial_v \otimes (vW + k_B T \partial_v W / m) \qquad (12)$$

where $I$ is the unit tensor. The equilibrium solution of Eq. (12) is the Maxwell-Boltzmann distribution density. Neglecting the first inertial term in Eq. (11) at large friction leads to an expression $\dot{R} = (m\gamma_0 I + \tau_0 \nabla \otimes \nabla U)^{-1} \cdot (F - \nabla U)$ for the particle velocity. Hence, the corresponding Smoluchowski equation for the probability density $\rho(r,t)$ acquires in the classical case the form

$$\partial_t \rho = \nabla \cdot [(m\gamma_0 I + \tau_0 \nabla \otimes \nabla U)^{-1} \cdot (\rho \nabla U + k_B T \nabla \rho)] \qquad (13)$$

Its equilibrium solution is the Boltzmann distribution. Since $\rho = \int_{-\infty}^{\infty} W dv$, Eq. (13) can be derived directly from Eq. (12) in standard ways [9]. To examine the correctness of the equations above, let us consider again the harmonic oscillator with $U = m\omega_0^2 r^2 / 2$. In this case, Eq. (12) reduces to the usual Klein-Kramers equation

$$\partial_t W + v \cdot \nabla W - \omega_0^2 r \cdot \partial_v W = \gamma_{\omega_0} \partial_v \cdot (vW + k_B T \partial_v W / m) \qquad (14)$$

while Eq. (13) becomes the usual Smoluchowski equation

$$\partial_t \rho = \nabla \cdot (\omega_0^2 r \rho + k_B T \nabla \rho / m) / \gamma_{\omega_0} \qquad (15)$$

The equations above correlate well to Eq. (5). The solution of Eq. (15) is a Gaussian distribution density with dispersion given by $\sigma^2 = (k_B T / m\omega_0^2)[1 - \exp(-2\omega_0^2 t / \gamma_{\omega_0})]$, which reduces in the case of vacuum ($\gamma_0 \equiv 0$) to $\sigma^2 = (k_B T / m\omega_0^2)[1 - \exp(-2t / \tau_0)]$. The latter indicates an extremely fast $\omega_0$-independent relaxation to the thermal equilibrium with $\sigma_{eq}^2 = k_B T / m\omega_0^2$.

In the Soviet school [10] a generalized Klein-Kramers equation is derived via the Bogolyubov method for a classical colored Langevin force

$$\partial_t W + v \cdot \nabla W - \nabla U \cdot \partial_v W / m = \partial_v \cdot \int_{-\infty}^{0} \frac{C_{FF}(\tau)}{mk_B T} \cdot (vW + k_B T \partial_v W / m)_{t+\tau} d\tau \qquad (16)$$

This equation is a modification of the Nakajima-Zwanzig master equation in the Born approximation [6]. Substituting the particular expression $C_{FF} = 2mk_B T(\gamma_0 - \tau_0 \partial_\tau^2)\delta(\tau)I$ for the Langevin force autocorrelation function and integrating on time yield a new linear differential Klein-Kramers equation

$$\partial_t W + v \cdot \nabla W - \nabla U \cdot \partial_v W / m = (\gamma_0 - \tau_0 \partial_t^2)\partial_v \cdot (vW + k_B T \partial_v W / m) \qquad (17)$$

Equation (17) is not Markovian due to the second derivative on time, reflecting the Abraham-Lorentz force. Its equilibrium solution is the Maxwell-Boltzmann distribution. The Fourier transform of Eq. (17) is a natural generalization of the classical Klein-Kramers equation for the case of an arbitrary frequency-dependent friction coefficient. One can derive after standard integrations [9] the corresponding generalization of the Smoluchowski equation

$$\partial_t^2 \rho + \gamma_0 \partial_t \rho = \nabla \cdot (\rho \nabla U + k_B T \nabla \rho) / m + \tau_0 \partial_t^3 \rho \qquad (18)$$

For a harmonic oscillator one can easily derive by integration of Eq. (18) that the evolution of the position dispersion obeys the following equation $\partial_t^2 \sigma^2 + \gamma_0 \partial_t \sigma^2 + 2\omega_0^2 \sigma^2 = \tau_0 \partial_t^3 \sigma^2 + 2k_B T / m$. Although the latter is derived in the classical case, one can add a posteriorly the quantum particle effect via the Heisenberg velocity dispersion [8] to obtain

$$\partial_t^2 \sigma^2 + \gamma_0 \partial_t \sigma^2 + 2\omega_0^2 \sigma^2 = \tau_0 \partial_t^3 \sigma^2 + 2k_B T / m + 2(\hbar / 2m\sigma)^2 \qquad (19)$$

For a free particle in vacuum at zero temperature this equation can be simplified at high friction to $-\tau_0 \partial_t^3 \sigma^2 = \hbar^2 / 2m^2 \sigma^2$. Its solution $\sigma^2 = (6\hbar\tau_0 / m)(t / 3\tau_0)^{3/2}$ describes super-diffusive Brownian motion of a quantum particle in classical environment and, for this reason, it differs from previously derived sub-diffusive expression $\sigma^2 = (2\hbar\tau_0 / \pi m) \ln(t / \tau_0)$ for a classical particle moving in quantum vacuum [8].

## Appendix

The second fluctuation-dissipation theorem $S_{FF} = m\gamma_\omega \hbar\omega \coth(\hbar\omega / 2k_B T)$ expresses the spectral density of the Langevin force by the total friction coefficient. To find their particular forms, however, more information is required from the liner response for instance. Hereafter, some general models are represented, which could discriminate some important cases. The most famous is the Ohmic friction, where the friction coefficient $\gamma_\omega = \gamma_0$ is frequency independent. In this case the spectral density of the Langevin force reads $S_{FF} = m\gamma_0 \hbar\omega \coth(\hbar\omega / 2k_B T)$, which corresponds to the autocorrelation function $C_{FF} = m\hbar\gamma_0 \cot(\hbar\partial_\tau / 2k_B T) \partial_\tau \delta(\tau)$. Another very important model is the white noise with a constant spectral density $S_{FF} = 2mk_B T\gamma_0$ and delta autocorrelation function $C_{FF} = 2mk_B T\gamma_0 \delta(\tau)$. According to the fluctuation-dissipation theorem it corresponds to the following friction coefficient $\gamma_\omega = \gamma_0 \tanh(\hbar\omega / 2k_B T) / (\hbar\omega / 2k_B T)$. These two models coincide at low frequency $\omega < 2k_B T / \hbar$, which is always the case in the classical limit $\hbar \to 0$. Since fluctuation and dissipation are complimentary processes, we are tempted to define twins with identical asymptotic behavior at high frequency, i.e. $S_{FF}(\omega \to \infty) / S_{FF}(0) = \gamma_{\omega \to \infty} / \gamma_0$. A Langevin force with spectral density $S_{FF} = 2mk_B T\gamma_0 \cosh(\hbar\omega / 2k_B T)$ and autocorrelation function $C_{FF} = 2mk_B T\gamma_0 \cos(\hbar\partial_\tau / 2k_B T) \delta(\tau)$ is such a twin process. The reciprocal value of the corresponding friction coefficient $2k_B T\gamma_0 / \gamma_\omega = \hbar\omega / \sinh(\hbar\omega / 2k_B T)$ is proportional to the harmonic oscillator partition function and thus it correlates well to a kinetic constant from the transition state theory [11]. Note, that expansion the friction coefficient at low frequency

$$\gamma_\omega = \gamma_0 + \gamma_0 (\hbar\omega / 2k_B T)^2 / 6 \qquad (20)$$

resembles the radiative force if $\gamma_0 = 24(k_B T / \hbar)^2 \tau_0$. This raises a question about the existence of friction in vacuum at finite temperature. Another twin example possesses a friction coefficient $\gamma_\omega = \gamma_0 / \cosh(\hbar\omega / 2k_B T)$ and Langevin force spectral density $S_{FF} = m\gamma_0 \hbar\omega / \sinh(\hbar\omega / 2k_B T)$ and autocorrelation function $C_{FF} = m\hbar\gamma_0 \sin(\hbar\partial_\tau / 2k_B T)^{-1} \partial_\tau \delta(\tau)$. This model seems the most appropriate one, since both $S_{FF}$ and $\gamma_\omega$ vanish naturally at infinite frequency as physically required.

Another interesting question is what kind friction corresponds to maximal spectral density of the particle velocity. Obviously, Eq. (7) possesses a maximum in respect to the friction coefficient, since at large friction the particle cannot move quickly, while at low friction the fluctuations are weak. The maximum $S_{VV} = \hbar \coth(\hbar\omega/2k_B T)/2m$ corresponds to $\gamma_\omega \equiv \omega$ and reduces to $S_{VV} = \hbar/2m$ for a white noise at zero temperature. In the classical limit $S_{VV} = k_B T/m\omega$ is one for Flicker noise. The Langevin force spectral densities read $S_{FF} = m\hbar\omega^2 \coth(\hbar\omega/2k_B T)$, $S_{FF} = m\hbar\omega^2$ and $S_{FF} = 2mk_B T\omega$, respectively. The latter resembles the Langevin force spectral density $S_{FF} = m\gamma_0 \hbar\omega$ for the quantum Brownian motion at low temperature if $\gamma_0 = 2k_B T/\hbar$.

[1] W. T. Grandy Jr., *Introduction to Electrodynamics and Radiation*, Academic Press, New York, 1970

[2] L. de la Peña and A. M. Cetto, *The Quantum Dice: An Introduction to Stochastic Electrodynamics*, Kluwer, Dordrecht, 1996

[3] L. de la Peña, A. M. Cetto and A. Valdes-Hernandez, *The Emerging Quantum: The Physics Behind Quantum Mechanics*, Springer, Heidelberg, 2015

[4] H. E. Puthoff, *Phys. Rev. D* **35** (1987) 3266

[5] P. R. Johnson and B. L. Hu, *Phys. Rev. D* **65** (2002) 065015

[6] U. Weiss, *Quantum Dissipative Systems*, World Scientific Publishing, Singapore, 2012

[7] R. Tsekov and E. Ruckenstein, *J. Chem. Phys.* **100** (1994) 1450

[8] R. Tsekov, *Chin. Phys. Lett.* **29** (2012) 120504

[9] H. Risken, *The Fokker-Planck Equation: Methods of Solution and Applications*, Springer, Heidelberg, 1989

[10] I. P. Bazarov, E. V. Gevorkyan and P. N. Nikolaev, *Non-equilibrium Thermodynamics and Physical Kinetics*, Moscow University Press, Moscow, 1989, p. 59

[11] R. Tsekov, *Int. J. Mol. Sci.* **2** (2001) 66